\documentclass{article}

\def\xxinput#1{\input#1}

\xxinput{vsolj02.sty}

\usepackage[dvipdfmx]{graphicx}
\usepackage[comma,colon]{natbib}

\usepackage[OT2,T1]{fontenc}

\def\cite{\citealt}
\setcitestyle{aysep={}}

\def\commenta{$^*$}
\def\commentb{$^\dagger$}

\newcounter{author}
\def\altaffilmark#1{$^{#1}$}
\def\altaffiltext#1{$^{#1}$\,}

\def\authorcount#1#2{{\refstepcounter{author}\label{#1}
                     \altaffiltext{\ref{#1}}{#2}}}

\begin{document}

\begin{center}

\title{Analysis of TESS observations of V844 Her during the 2020 superoutburst}

\author{
        Taichi~Kato\altaffilmark{\ref{affil:Kyoto}}
}
\email{tkato@kusastro.kyoto-u.ac.jp}

\authorcount{affil:Kyoto}{
     Department of Astronomy, Kyoto University, Sakyo-ku,
     Kyoto 606-8502, Japan}

\end{center}

\begin{abstract}
\xxinput{abst.inc}
\end{abstract}

   Superhump stages A, B and C in SU UMa-type (including
WZ Sge-type) dwarf novae were established in \citet{Pdot}
[for general information of cataclysmic variables and dwarf novae,
see e.g. \citet{war95book}].
In \citet{Pdot}, a figure representing stage A--C
was shown based on the 2000 ground-based observation of SW UMa
(in their figure 3), and this figure or its revision
(figure 1 in \cite{kat13qfromstageA})
has been used in various papers to explain the concept
of superhump stages.  The data for SW UMa,
however, were rather old and there were gaps in
the observations.  Although a figure based on continuous
satellite observations had been desired, only long cadence (LC)
Kepler data of V585 Lyr could be used nine years ago
\citep{kat13j1939v585lyrv516lyr}.  The Kepler LC data
were insufficient to draw an $O-C$ diagram directly
and \citet{kat13j1939v585lyrv516lyr} had to reconstruct
the original light curve by introducing a Bayesian method.

\begin{figure*}
\begin{center}
\includegraphics[width=16cm]{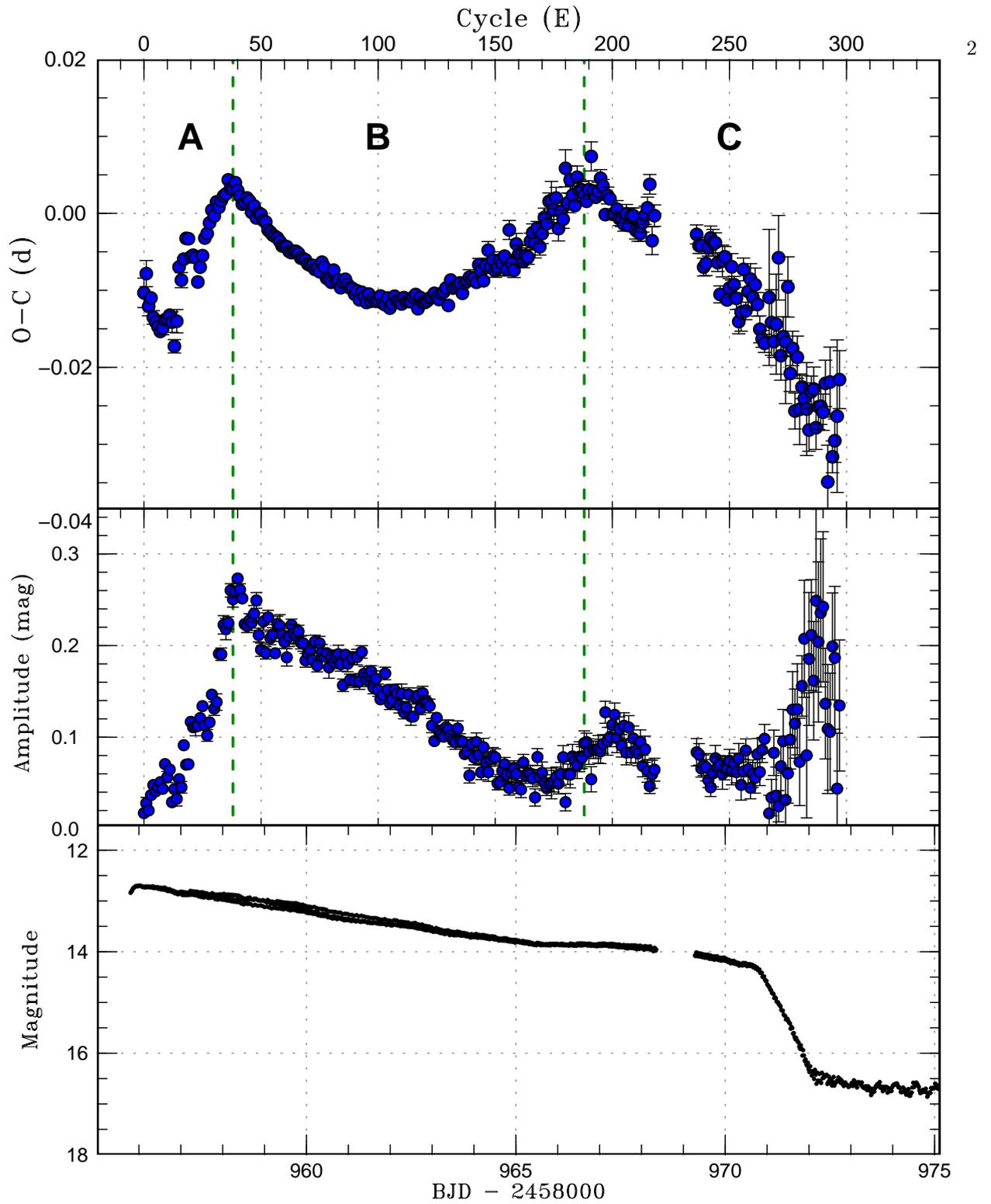}
\caption{
  Superhump stages in V844 Her during the 2020 superoutburst.
  (Upper:) $O-C$ variation.  The ephemeris of
  BJD(max) = 2458957.350$+$0.05592$E$ was used.
  The data are in table \ref{tab:shmax}.
  (Middle:) Superhump amplitude.  The amplitudes grew during
  stage A (textbook behavior of a short-period system).
  They decreased during stage B, and showed a slight growth
  when stage C started.
  (Lower:) TESS light curve.
  The data were binned to 0.019~d.  Slight brightening in
  stage C corresponds to the slight growth of superhumps
  in the middle panel.
}
\label{fig:humpall}
\end{center}
\end{figure*}

   The aim of this paper is to provide a corresponding figure
based on modern data which is equivalent to figure 3 in \citet{Pdot}.
I used Transiting Exoplanet Survey Satellite (TESS)
observations \citep{ric15TESS}\footnote{
  $<$https://tess.mit.edu/observations/$>$.
  The full light-curve
  is available at the Mikulski Archive for Space Telescope
  (MAST, $<$http://archive.stsci.edu/$>$).
} of the 2020 superoutburst of V844 Her, which is
an SU UMa-type dwarf nova very similar to SW UMa
\citep{ant96newvar,kat00v844her,oiz07v844her}
and has a similar short orbital period of 0.054643~d
\citep{tho02gwlibv844herdiuma}.
In addition to the greatly improved coverage and
statistics, the figures in this paper are provided
under the Creative Commons licence (CC-BY-NC) and
can be used without consideration of the copyright
of the publisher.  First of all, I show figure \ref{fig:humpall}.
This figure clearly demonstrate the ``textbook'' example of
superhump stages in a short-period SU UMa-type dwarf nova.
I describe how I obtained this figure.

\begin{figure*}
  \begin{center}
    \includegraphics[width=16cm]{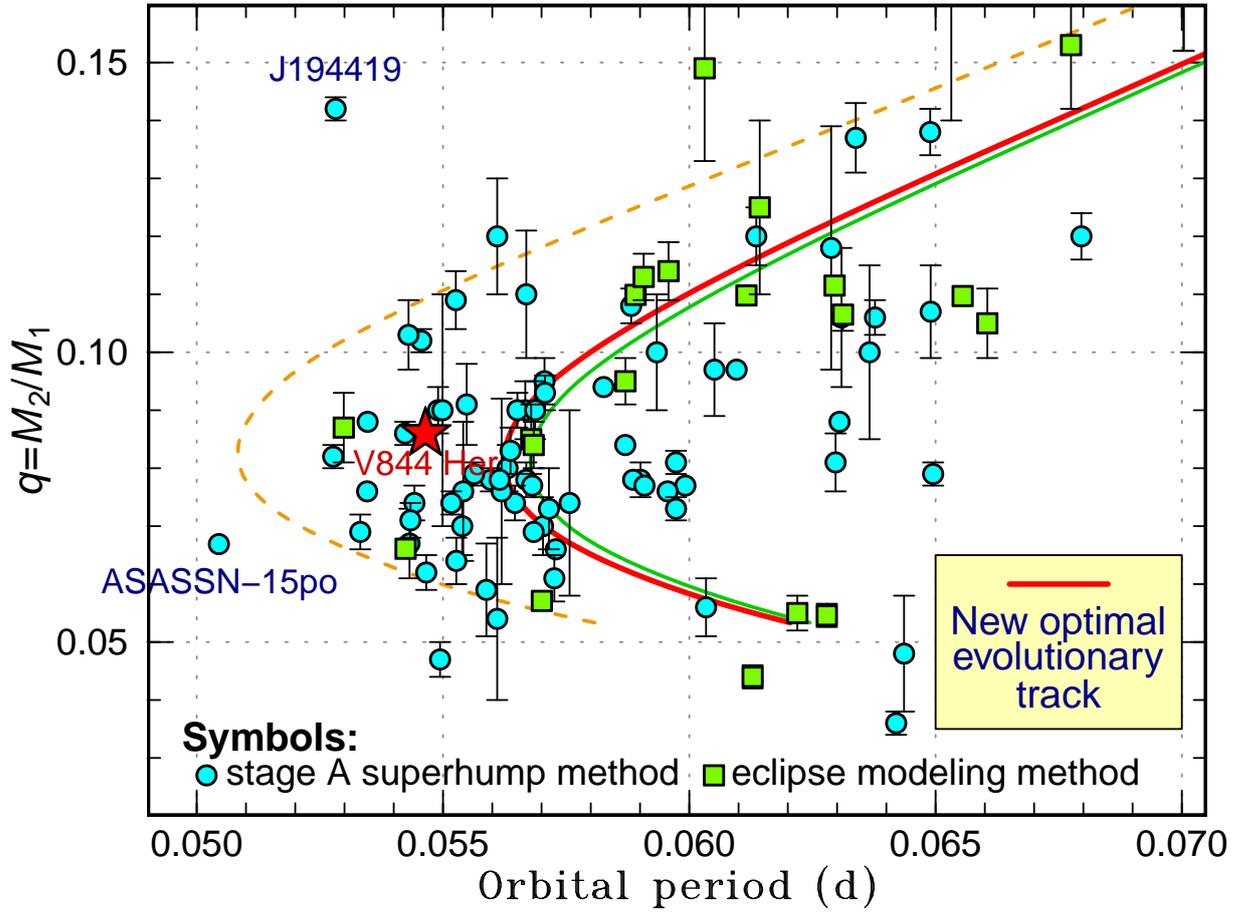}
  \end{center}
  \caption{Location of V844 Her (filled red star)
  on the relation between mass ratios ($q$) and
  orbital periods ($P_{\rm orb}$)
  determined by the eclipse modeling method and the stage A superhump
  method, enlargement around the period minimum.
  The dashed and green solid curves represent the standard and optimal
  evolutionary tracks in \citet{kni11CVdonor}, respectively.
  The revised optimal evolutionary track in \citet{kat22stageA}
  is shown by a red solid curve.
  See explanation in \citet{kat22stageA} for more details.
  }
  \label{fig:qall}
\end{figure*}

\begin{figure*}
\begin{center}
\includegraphics[width=16cm]{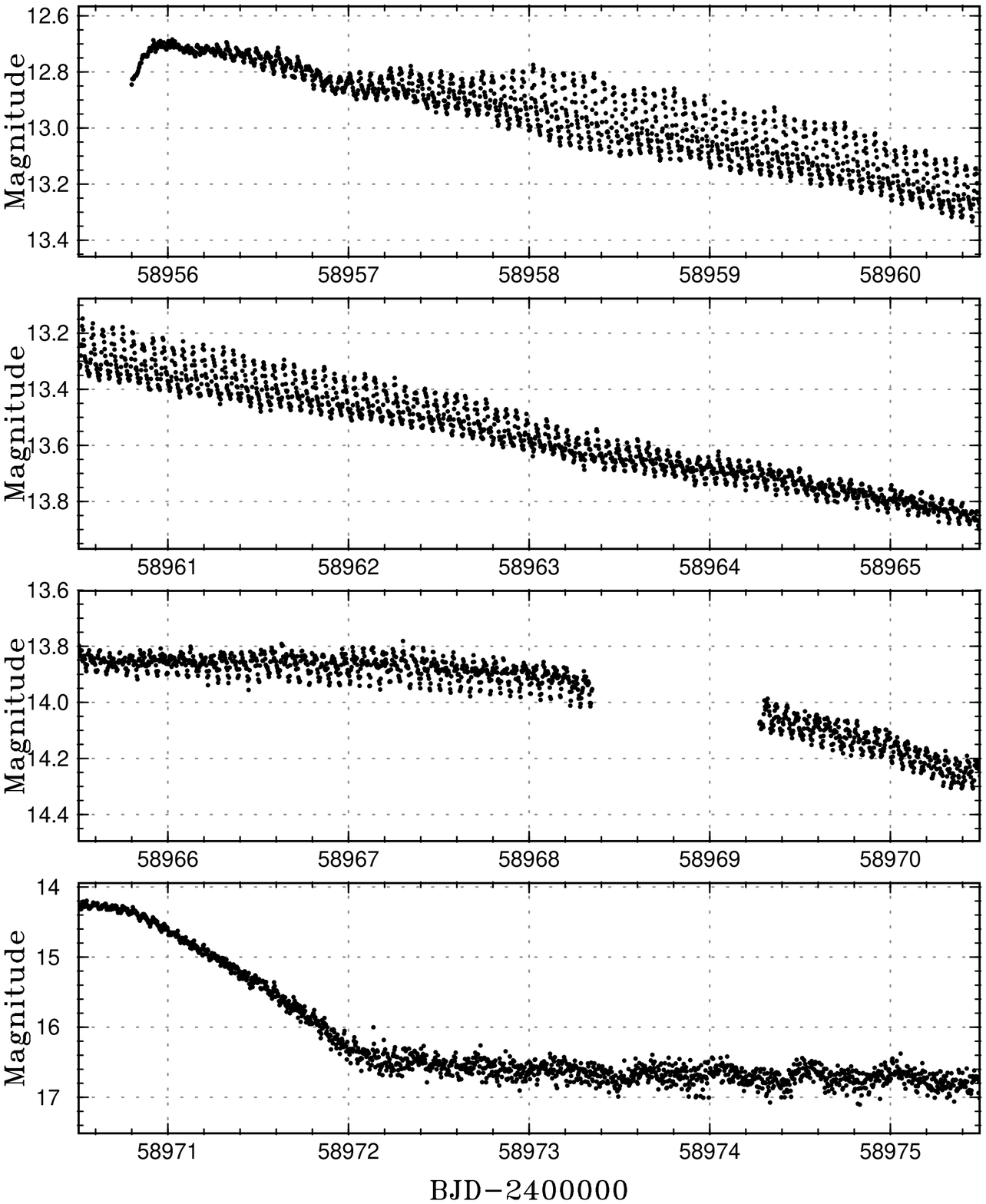}
\caption{
  TESS light curve of V844 Her during the 2020 superoutburst
  and early post-superoutburst phase.
  The magnitudes were defined as
  $-2.5\log_{10}({\rm flux}/15000)+10$.
  The data were binned to 0.003~d.
  Small wiggles with a period of $\sim$0.5~d during
  the post-superoutburst phase may be the same phenomenon
  of ``mini-rebrightenings'' recorded in the Kepler data
  of V585 Lyr \citep{kat13j1939v585lyrv516lyr}.
}
\label{fig:tess1}
\end{center}
\end{figure*}

   The TESS data of V844 Her during the 2020 superoutburst
and early post-superoutburst phase are shown in
figure \ref{fig:tess1}.  The superhump maxima were
obtained by the template fitting method introduced
in \citet{Pdot} and the quality of the $O-C$ values is
the same as in the series of ``Pdot papers''
\citep{Pdot,Pdot2,Pdot3,Pdot4,Pdot5,Pdot6,Pdot7,Pdot8,Pdot9,Pdot10}
for a collection of SU UMa/WZ Sge stars and
which were homogeneously determined using the same method.
This method has an advantage of much higher signal-to-noise and
thereby higher precision in determining maxima than
picking up the peaks \citep{Pdot}.
The representative superhump periods in different superhump stages
are given in table \ref{tab:shper}.
For stages A, B and C, the mean period ($P$) and the period
derivative ($P_{\rm dot} = \dot{P}/P$) are given.
For the post-superoutburst stage, individual maxima could
not be obtained by the template fitting method and
the periods were determined using
the Phase Dispersion Minimization (PDM, \cite{PDM})
method after removing long-term trends by locally-weighted
polynomial regression (LOWESS: \cite{LOWESS}).
The errors of periods by the PDM method were
estimated by the methods of \citet{fer89error} and \citet{Pdot2}.
I used 3-d windows considering the beat period between
$P_{\rm orb}$ and the superhump period ($P_{\rm SH}$).
Using $P_{\rm orb}$ and the period of stage A superhumps,
the fractional superhump excess in frequency
$\epsilon^* = 1-P_{\rm orb}/P_{\rm SH}$ for stage A is
0.0316(7).  This value corresponds to a mass ratio of
$q$=0.086(2) using the stage A superhump method
\citep{kat13qfromstageA,kat22stageA}.  The location
on the $P_{\rm orb}$--$q$ plane is shown in figure
\ref{fig:qall}, on which V844 Her was plotted on figure 11
in \citet{kat22stageA}.

\begin{figure*}
\begin{center}
\includegraphics[width=16cm]{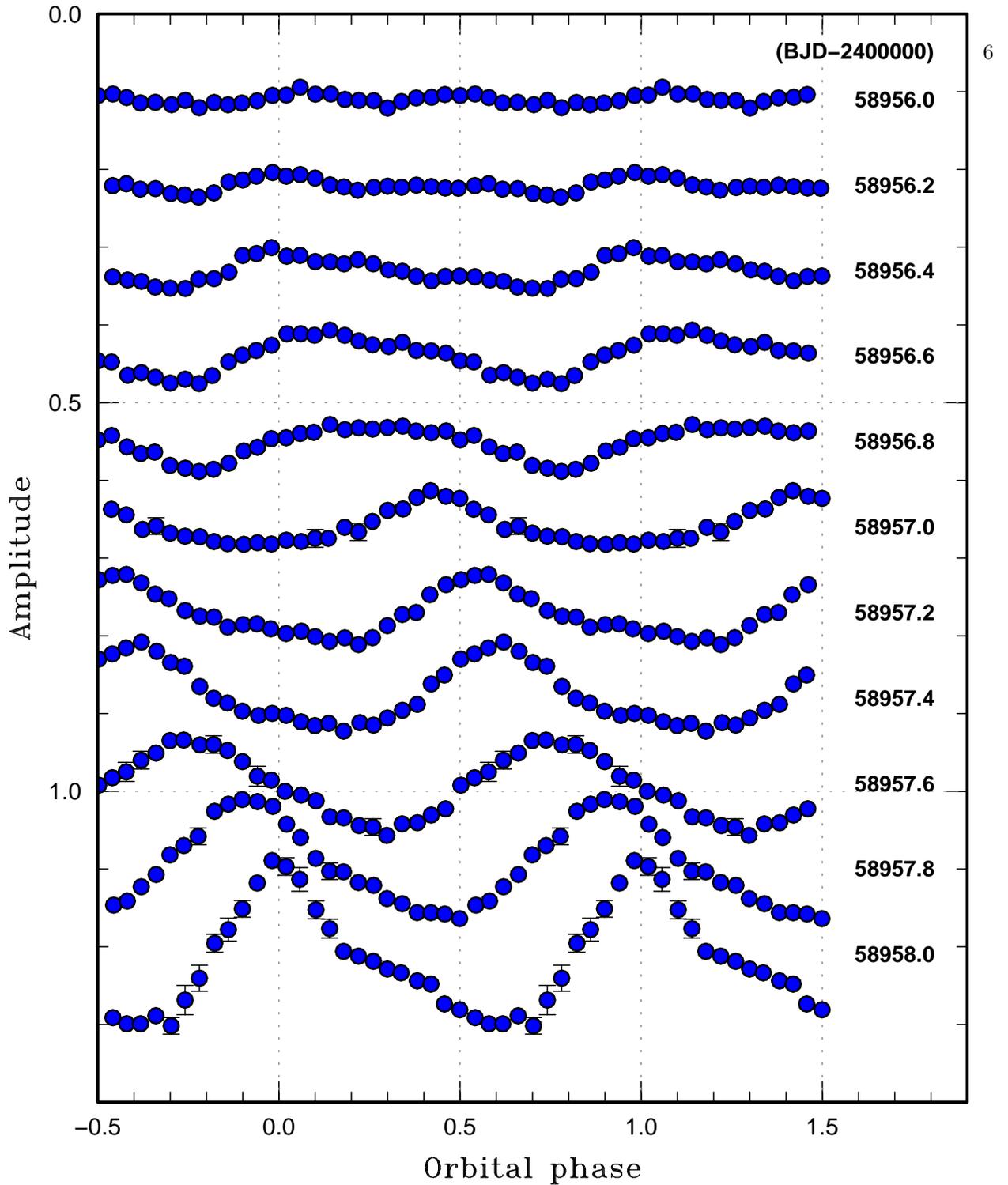}
\caption{
  Variation of profiles of superhumps in the growing phase.
  The zero phase and orbital period were defined as BJD 2458956.000
  (arbitrarily chosen) and 0.054643~d.
  0.2-d segments were used whose centers are shown on the right side
  of the figure.
  After BJD 2458957.0, superhumps (stage A superhumps)
  with a period longer than the orbital period developed.
  Between BJD 2458956.0 and 2458957.0, the period was closer
  to the orbital one, while the average epoch was on the smooth
  extension of stage A superhumps (see figure \ref{fig:humpall}).
}
\label{fig:prof}
\end{center}
\end{figure*}

\begin{figure*}
\begin{center}
\includegraphics[width=16cm]{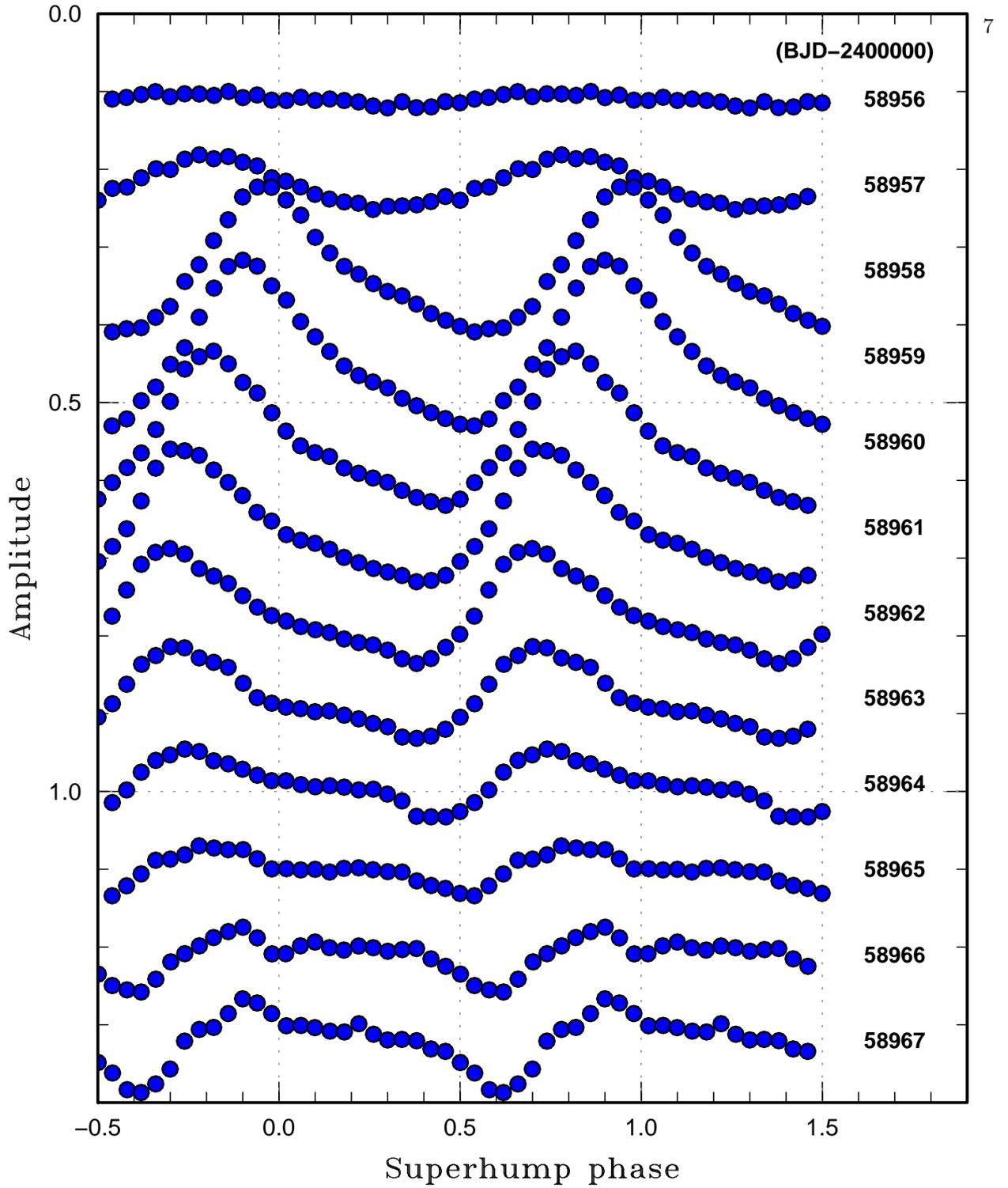}
\caption{
  Variation of profiles of superhumps in stages A and B.
  The zero phase and superhump period were defined as
  BJD 2458958.2479 and 0.055592~d.
  1-d segments were used whose centers are shown on the right side
  of the figure.  Systematic variation of the peak phase
  corresponds to the $O-C$ variation or period change.
}
\label{fig:prof2}
\end{center}
\end{figure*}

\begin{figure*}
\begin{center}
\includegraphics[width=16cm]{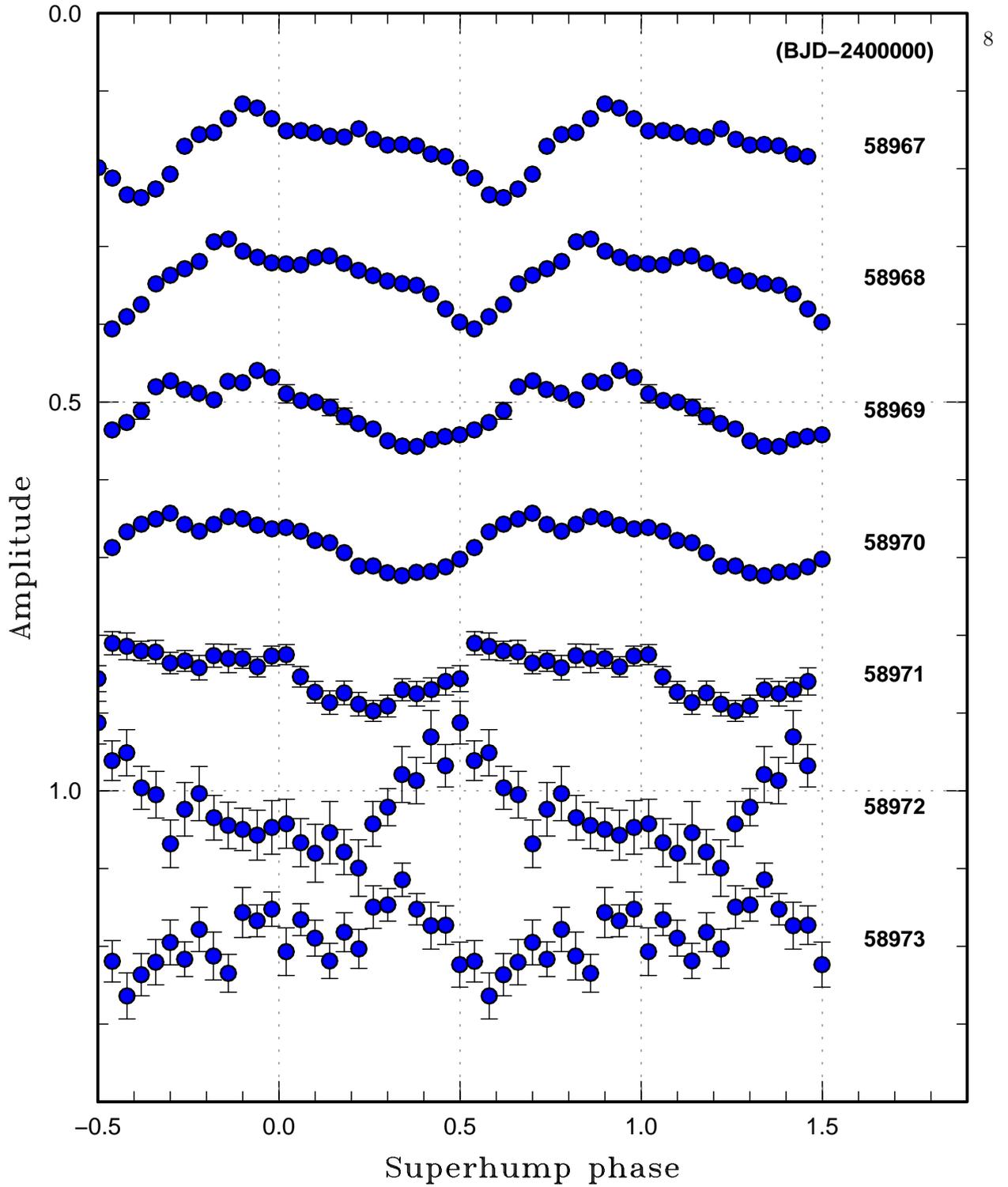}
\caption{
  Variation of profiles of superhumps from stage B-C
  transition to the early post-superoutburst phase.
  The zero phase and superhump period were defined as
  BJD 2458958.2479 and 0.055592~d.
  1-d segments were used whose centers are shown on the right side
  of the figure.  Although post-superoutburst superhumps
  were visible (after BJD 2458972), they became inapparent
  due to flickering.
}
\label{fig:prof3}
\end{center}
\end{figure*}

   The variation of the superhump profile is shown in
figures \ref{fig:prof}, \ref{fig:prof2} and \ref{fig:prof3}.
The most important finding in this paper is in the development
of superhumps in stage A (figure \ref{fig:prof}).
Although stage A with growing amplitudes is clearly present
in figure \ref{fig:humpall}, the $O-C$ variation is not
on a straight line.  The reason has been clarified in
figure \ref{fig:prof}: before stage A superhump started
to develop fully, there was a phase when the period
was close to (but slightly longer than) the orbital period.
The average epoch of humps in this phase, however, was
on the smooth extension of stage A superhumps
(see figure \ref{fig:humpall}).  It was not clear whether this
feature is special to V844 Her or it is a general phenomenon.
Observers should pay attention to the presence of this
phenomenon not to confuse this variation with early
superhumps seen in WZ Sge stars \citep{kat15wzsge}.
This is particularly true when the period is identified
only by frequency(period) analysis and no $O-C$ analysis
is performed.

   During the post-superoutburst phase (figure \ref{fig:prof3}),
superhumps rapidly decayed.  The superhump signal, however,
was detected with the same period by the PDM method
for 18~d in the post-superoutburst phase
(table \ref{tab:shper}; the gap between BJD 2458981
and 2458984 was due to the gap in TESS observation).
No orbital signal was detected during the post-superoutburst
phase as in the case of V585 Lyr.
The small wiggles with a period of $\sim$0.5~d during
the post-superoutburst phase (figure \ref{fig:tess1})
may be the same phenomenon of ``mini-rebrightenings''
recorded in the Kepler data of
V585 Lyr \citep{kat13j1939v585lyrv516lyr}.

   The times of maxima of superhumps used in this study
are listed in table \ref{tab:shmax}.

\begin{table*}
\caption{Superhump period in V844 Her}\label{tab:shper}
\begin{center}
\begin{tabular}{ccccc}
\hline
Stage & $E$ & BJD$-$2400000 & Period (d) & $P_{\rm dot}\times 10^5$ \\
\hline
A &   0--38  & 58956.11--58958.25 & 0.056425(43) & 44(13) \\
B &  38--188 & 58958.25--58966.64 & 0.055928(9) & 9.7(2) \\
C & 188--297 & 58966.64--58972.71 & 0.055638(12) & $-$9.0(12) \\
Post-superoutburst & -- & 58972--58975 & 0.05554(4) & -- \\
Post-superoutburst & -- & 58975--58978 & 0.05554(4) & -- \\
Post-superoutburst & -- & 58978--58981 & 0.05554(4) & -- \\
Post-superoutburst & -- & 58984--58987 & 0.05553(6) & -- \\
Post-superoutburst & -- & 58987--58990 & 0.05558(8) & -- \\
\hline
\end{tabular}
\end{center}
\end{table*}

\begin{table*}
\caption{Times of superhump maxima in V844 Her}\label{tab:shmax}
\begin{center}
\begin{tabular}{cccc|cccc|cccc}
\hline
$E$ & $T$\commenta & error & amp\commentb &
$E$ & $T$ & error & amp &
$E$ & $T$ & error & amp \\
\hline
\xxinput{tab1-1.inc}
\hline
\multicolumn{12}{l}{\commenta BJD$-$2458000.} \\
\multicolumn{12}{l}{\commentb Amplitude (mag).} \\
\end{tabular}
\end{center}
\end{table*}

\addtocounter{table}{-1}

\begin{table*}
\caption{Times of superhump maxima in V844 Her (continued)}
\begin{center}
\begin{tabular}{cccc|cccc|cccc}
\hline
$E$ & $T$\commenta & error & amp\commentb &
$E$ & $T$ & error & amp &
$E$ & $T$ & error & amp \\
\hline
\xxinput{tab1-2.inc}
\hline
\multicolumn{12}{l}{\commenta BJD$-$2458000.} \\
\multicolumn{12}{l}{\commentb Amplitude (mag).} \\
\end{tabular}
\end{center}
\end{table*}

\section*{Acknowledgements}

This work was supported by JSPS KAKENHI Grant Number 21K03616.
The author is grateful to the TESS team for making their
data available to the public.
I am grateful to Naoto Kojiguchi for helping downloading
the TESS data.

\section*{List of objects in this paper}
\xxinput{objlist.inc}

\section*{References}

We provide two forms of the references section (for ADS
and as published) so that the references can be easily
incorporated into ADS.

\renewcommand\refname{\textbf{References (for ADS)}}

\newcommand{\noop}[1]{}\newcommand{\hyphalt}{-}

\xxinput{v844tessaph.bbl}

\renewcommand\refname{\textbf{References (as published)}}
\xxinput{v844tess.bbl.vsolj}

\end{document}